\documentclass[prl,showpacs,showkeys,twocolumn]{revtex4}
\usepackage{bm}
\usepackage{amssymb,amsmath,amsthm}
\usepackage{mathrsfs}
\usepackage{daytime}
\usepackage{dayofweek}

\begin{document}
\title{Bounding the greybody factors for Schwarzschild black holes}
\author{Petarpa Boonserm}
\email{petarpa.boonserm@mcs.vuw.ac.nz}
\affiliation{School of Mathematics, Statistics, and Computer Science, 
Victoria University of Wellington, PO Box 600, Wellington, New Zealand}
\author{Matt Visser}
\email{matt.visser@mcs.vuw.ac.nz}
\affiliation{School of Mathematics, Statistics, and Computer Science, 
Victoria University of Wellington, PO Box 600, Wellington, New Zealand}

\date{13 June 2008; 
\LaTeX-ed \DayOfWeek, \today; \daytime }
\begin{abstract}
Greybody factors in black hole physics modify the naive Planckian spectrum that is predicted for Hawking radiation when working in the limit of geometrical optics.  We   consider the Schwarzschild geometry in (3+1) dimensions, and analyze the Regge--Wheeler equation for arbitrary particle spin $s$ and wave-mode angular momentum $\ell$, deriving rigourous bounds on the greybody factors as a function of $s$, $\ell$, wave frequency $\omega$, and the black hole mass $m$.

\end{abstract}


\maketitle

\newcommand{\scri}{\mathscr{I}}
\newcommand{\sun}{\ensuremath{\odot}}
\def\d{{\mathrm{d}}}
\def\tr{{\mathrm{tr}}}
\def\sech{{\mathrm{sech}}}
\def\etc{\emph{etc}}
\def\ie{{\emph{i.e.}}}
\def\implies{\Rightarrow}
\newcommand{\pacomment}[1]{ {\em \color{blue} #1}}

\def\lint{\hbox{\Large $\displaystyle\int$}} 
\def\hint{\hbox{\Huge $\displaystyle\int$}}  
\def\Re{{\mathrm{Re}}}
\def\Im{{\mathrm{Im}}}

\section{Introduction}

Black-hole greybody factors modify the spectrum of Hawking radiation seen at spatial infinity~\cite{Hawking}, so that it is not quite Planckian~\cite{Page}. 
There is a vast scientific literature dealing with estimates of these black-hole greybody factors, using a wide variety of techniques~\cite{greybody}. 

Unfortunately, most of these calculations adopt various approximations that move one away from the physically most important regions of parameter space.  Sometimes one is forced into the extremal limit, sometimes one is forced to asymptotically high or low frequencies, sometimes techniques work only away from (3+1) dimensions, sometimes the nature of the approximation is uncontrolled. As a specific example, \emph{monodromy} techniques fail for $s=1$ (photons)~\cite{monodromy}, which is observationally one of the most important cases one would wish to consider. 

Faced with these limitations, we ask a slightly different question: Restricting attention to the physically most important situations (Schwarzschild black holes, (3+1) dimensions, intermediate frequencies, unconstrained spin and angular momentum) is it possible to at least place rigorous (and hopefully simple) analytic \emph{bounds} on the greybody factors? 

By considering the Regge--Wheeler equation for excitations around Schwarzschild spacetime, and adapting the general analysis of references~\cite{bounds1, bounds2}, we shall demonstrate that rigorous analytic bounds are indeed achievable. While these bounds may not answer all the physical questions one might legitimately wish to ask, they are a solid step in the right direction. 

\section{Regge--Wheeler equation}

In terms of the tortoise coordinate $r_*$ the Regge--Wheeler equation ($G_N\to1$) is 
\begin{equation}
{d^2 \psi\over d r_*^2} =   [ \omega^2 - V(r) ] \psi,
\end{equation}
where for the specific case of a Schwarzschild black hole
\begin{equation}
{dr\over dr_*} = 1-{2m\over r},
\end{equation}
and the Regge--Wheeler potential is
\begin{equation}
V(r) = \left(1-{2m\over r}\right) \left[ {\ell(\ell+1)\over r^2} + {2m(1-s^2)\over r^3} \right].
\end{equation}
Here $s$ is the spin of the particle and $\ell$ is the angular momentum of the specific wave mode under consideration, with $\ell\geq s$. Thus $V(r)\geq0$ outside the horizon, where $r\in(2m,\infty)$. The greybody factors we are interested in are just the transmission probabilities for wave modes propagating through this Regge--Wheeler potential.

\begin{itemize}
\item 

Despite comments often encountered in the literature, one \emph{can} explicitly solve for $r$ as a function of the tortoise coordinate $r_*$ --- in terms of Lambert $W$ functions we have
\begin{equation}
r(r_*) = 2m\left[ 1 + W( e^{[r_*-2m]/2m}) \right],
\end{equation}
whereas
\begin{equation}
r_*(r) = r + 2m\ln\left[{r-2m\over2m}\right].
\end{equation}
Unfortunately this formal result is less useful than one might suppose.

\item
Despite other comments often encountered in the literature, one can also explicitly solve the Regge--Wheeler equation --- now in terms of Heun functions~\cite{Heun}. Unfortunately this is again less useful than one might suppose, this time because relatively little is known about the analytical behaviour of Heun functions --- this is an area of ongoing research in mathematical analysis~\cite{Heun2}.
\end{itemize}

\section{Bounds}

The general bounds developed in references~\cite{bounds1, bounds2} can, in the current situation, be written as
\begin{equation}
T \geq \sech^2\left\{ \int_{-\infty}^{\infty}  \vartheta \; \d r_* \right\}.
\end{equation}
Here $T$ is the transmission probability (greybody factor), and $\vartheta$ is the function
\begin{equation}
\vartheta = {\sqrt{ (h')^2 + [\omega^2-V- h^2]^2}\over2 h }.
\label{E:b0}
\end{equation}
Furthermore, $h$ is some positive function, $h(r_*)>0$, satisfying the limits $h(-\infty)=h(+\infty)=\omega$, which is otherwise arbitrary. Two different derivations of this general result, and numerous consistency checks, can be found in  references~\cite{bounds1, bounds2}.  

(These bounds were originally developed as a technical step when studying the completely unrelated issue of sonoluminescence~\cite{sonoluminescence},  and since then have also been used to place limits on particle production in analogue spacetimes~\cite{analogue} and resonant cavities~\cite{cavity}, to investigate qubit master equations~\cite{qubit}, and to motivate further general investigations of one-dimensional scattering theory~\cite{one-dim}.)  
For current purposes, the most useful practical results are obtained by considering two special cases: 

(1) If we set $h=\omega$ then
\begin{equation}
T \geq \sech^2\left\{ {1\over2\omega} \int_{-\infty}^{\infty} V(r_*) \; \d r_* \right\},
\end{equation}
whence
\begin{equation}
T \geq \sech^2\left\{ {1\over2\omega} \int_{2m}^{\infty}  
\left[ {\ell(\ell+1)\over r^2} + {2m(1-s^2)\over r^3} \right] \; \d r \right\}.
\end{equation}
Therefore, since the remaining integral is trivial, we obtain our first explicit bound:
\begin{equation}
T \geq \sech^2\left\{ {2\ell(\ell+1) +(1-s^2)\over 8\omega m} \right\}.
\label{E:b1}
\end{equation}
That is:
\begin{equation}
T \geq \sech^2\left\{ {(\ell+1)^2 +(\ell^2-s^2)\over 8\omega m} \right\}.
\label{E:b11}
\end{equation}
Note that this bound is meaningful for all frequencies. This is sufficient to tell us that at high frequencies the Regge--Wheeler  barrier is almost fully transparent, while even at arbitrarily low frequencies some nonzero fraction of the Hawking flux will tunnel through. A particularly nice feature of this first bound is that it is so easy to write down for arbitrary $s$ and $\ell$.

\medskip
(2) If we now set $h=\sqrt{\omega^2-V}$, which in this case implicitly means that we are not permitting any classically forbidden region, then
\begin{equation}
T \geq \sech^2\left\{ {1\over2} \int_{-\infty}^{\infty} \left|{h'\over h}\right| \; \d r_* \right\}.
\end{equation}
Since for arbitrary $s$ and $\ell$ the Regge--Wheeler potential is easily seen to have a unique peak at which it is a maximum, this becomes
\begin{eqnarray}
T &\geq& \sech^2\left\{ \ln\left( {h_\mathrm{peak}\over h_\infty } \right) \right\}
\\
&=& 
\sech^2\left\{ \ln\left( {\sqrt{\omega^2-V_\mathrm{peak}}\over \omega} \right) \right\},
\end{eqnarray}
which is easily seen to be monotonic decreasing as a function of $V_\mathrm{peak}$. However calculating the location of the peak, and value of the Regge--Wheeler potential at the peak is somewhat more tedious than evaluating the previous bound (\ref{E:b1}). Note that the present bound fails, and gives no useful information, once $\omega^2 < V_\mathrm{peak}$, corresponding to a classically forbidden region.  More explicitly, the bound can be rewritten as:
\begin{equation}
T \geq {4 \omega^2 (\omega^2-V_\mathrm{peak})\over (  2\omega^2 - V_\mathrm{peak})^2} = 
1 - {V_\mathrm{peak}^2\over (  2\omega^2 - V_\mathrm{peak})^2}.
\end{equation}

\bigskip
Let us now consider various sub-cases:
\begin{itemize}
\item 
For $s=1$ (ie, photons) the situation simplifies considerably. (Remember, this is the case for which monodromy techniques fail~\cite{monodromy}.) For $s=1$ we have $r_\mathrm{peak} = 3m$ and 
\begin{equation}
V_\mathrm{peak} = {\ell(\ell+1)\over27 m^2}.
\end{equation}
Consequently
\begin{equation}
T_{s=1} \geq {108 \omega^2 m^2 [27\omega^2 m^2-\ell(\ell+1)]\over [  54\omega^2 m^2 -\ell(\ell+1)]^2}.
\end{equation}
In almost the entire region where this bound applies ($\omega^2 > V_\mathrm{peak}$) it is in fact a better bound than (\ref{E:b1}) above.

\item
For $s=0$ (ie, scalars) and $\ell=0$ (the $s$-wave),   we have $r_\mathrm{peak} = 8m/3$ and 
\begin{equation}
V_\mathrm{peak} = {27\over1024 m^2}. 
\end{equation}
Consequently
\begin{equation}
T_{s=0,\ell=0} \geq {4096 \omega^2 m^2 [1024\omega^2 m^2-27]\over [2048\omega^2 m^2 -27]^2}.
\end{equation}
In a large fraction of the  region where this bound applies it is in fact a better bound than (\ref{E:b1}) above.

\item
For $s=0$ but $\ell\geq 1$ it is easy to see that throughout the black hole exterior, $\forall r \in(2m,\infty)$, we have 
\begin{equation}
V_{s=0,\ell\geq1}(r) < \left(1-{2m\over r}\right) \left[ {\ell^2+\ell+1\over r^2}\right],
\end{equation}
which is the $s=1$ potential with the replacement $\ell(\ell+1)\to \ell^2+\ell+1$. This bound on the potential has its maximum at $r_\mathrm{peak} = 3m$, implying
\begin{equation} 
V_{\mathrm{peak},s=0,\ell\geq1} <  {\ell^2+\ell+1\over 27 m^2}. 
\end{equation}
Therefore  the monotonicity of the bound on the greybody factor implies  
\begin{equation}
T_{s=0,\ell\geq1} >  {108 \omega^2 m^2 [27\omega^2 m^2-(\ell^2+\ell+1)]\over [  54\omega^2 m^2 -(\ell^2+\ell+1)]^2},
\end{equation}
(for $\omega$, $m$, and $\ell$ held fixed, and subject to $s\leq \ell$).

\item
For $s>1$ it is easy to see that throughout the black hole exterior, $\forall r \in(2m,\infty)$, keeping $\ell$ held fixed, we have $V_{s>1}(r) < V_{s=1}(r)$.  Therefore
\begin{equation} 
V_{\mathrm{peak},s>1} < V_{\mathrm{peak},s=1}. 
\end{equation}
Therefore  the monotonicity of the bound on the greybody factor implies  
\begin{equation}
T_{s>1} >   {108 \omega^2 m^2 [27\omega^2 m^2-\ell(\ell+1)]\over [  54\omega^2 m^2 -\ell(\ell+1)]^2},
\end{equation}
(for $\omega$, $m$, and $\ell$ held fixed, and subject to $s\leq \ell$).

\item
More generally, it useful to define
\begin{equation}
\epsilon= {1-s^2\over \ell(\ell+1)}.
\end{equation}
Excluding the case $(s,\ell)=(0,0)$, which was explicitly dealt with above, the remainder of the physically interesting region is confined to the range $\epsilon\in(-1,+1/2]$. Then a brief computation yields
\begin{equation}
r_\mathrm{peak} = 3m \left\{ 1 - {\epsilon\over 9} + \mathcal{O}(\epsilon^2)\right\},
\end{equation}
and 
\begin{equation}
V_\mathrm{peak} = {\ell(\ell+1)\over27 m^2}  \left\{ 1 +{2\epsilon\over3} + \mathcal{O}(\epsilon^2)\right\}.
\end{equation}
In fact one can show that
\begin{equation}
V_\mathrm{peak} <  {\ell(\ell+1)\over20 m^2}
\end{equation}
over the physically interesting range. (This bound on  $V_\mathrm{peak}$ is tightest for $(s,\ell) = (0,1)$, corresponding to $\epsilon=+1/2$, where it provides a better than 1\% estimate, and becomes progressively weaker as one moves to $\epsilon=-1$.)  This then implies
\begin{equation}
T_{(s,\ell)\neq(0,0)} > {80 \omega^2 m^2 [20\omega^2 m^2-\ell(\ell+1)]\over [  40\omega^2 m^2 -\ell(\ell+1)]^2}.
\end{equation}
As always there is a trade-off between strength of the bound and the ease with which it can be written down.
\end{itemize}
While this second set of bounds has required a little more case by case analysis, observe that this second set of bounds provides much stronger information at very high frequencies, where in fact
\begin{equation}
T \geq 1 - \mathcal{O}[V_\mathrm{peak} \;\omega^{-4}].
\end{equation}
Unfortunately this second set of bounds is (because of details in the derivation, see~\cite{bounds1,bounds2}) not capable of providing information once the frequency has dropped low enough for the problem to develop classical turning points --- in other words a problem with a classically forbidden region is not amenable to treatment using bounds of the second class considered above. For sufficently low frequencies, bounds of the form (\ref{E:b1}) are more appropriate, with
\begin{equation}
T \geq \mathcal{O}\left( \exp\{-1/\omega\}\right) . 
\end{equation}
What we have not done, at least not yet, is to use the full generality implicit in equation (\ref{E:b0}). Subject to rather mild constraints, there is a freely specifiable function $h(r_*)$ available that can potentially be used to extract tighter bounds. Work along these lines is continuing.

\section{Discussion}

The study of black hole greybody factors~\cite{greybody}, and (once one moves into the complex plane), the closely related problem of locating the quasinormal modes~\cite{monodromy,qnm,qnmv}, is a subject that has attracted a vast amount of interest. In the present article we have developed a complementary set of results --- we have sought and obtained several rigorous analytic bounds that can be placed on the greybody factors. While these bounds are not necessarily tight bounds on the exact greybody factors they do serve to focus attention on general and robust features of these greybody factors, and provide a new way of extracting physical information. For instance, in the current formalism, (as opposed to, for instance, monodromy techniques~\cite{monodromy}), it is manifestly clear that one does not have to know anything about what is going on inside the black hole in order to obtain information regarding the greybody factors. This is as it should be, since physically the greybody factors are simply transmission coefficients relating the horizon to spatial infinity, and make no intrinsic reference to the nature of the  central singularity. Looking further afield, here should be no intrinsic difficulty in extending these results to Reissner--Nordstr\"om black holes, dilaton black holes, or to higher dimensions --- all that is really needed is an exact expression for the Regge--Wheeler potential. Ultimately, it is perhaps more interesting to see if one can significantly improve these bounds in some qualitative manner, perhaps by making a more strategic choice for the essentially free function $h(r_*)$. 

\section*{Acknowledgments}

This research was supported by the Marsden Fund administered by the Royal Society of New Zealand. PB was additionally supported by a scholarship from the Royal Government of Thailand. 
MV wishes to specifically thank Eleftherios Papantonopoulos for his comments and questions,  and to thank the
Centro de Estudios Cient\'ificos (Valdivia, Chile) for hospitality.



\end{document}